\documentclass{article}

\usepackage{graphicx}  
\usepackage{amsmath}   
\usepackage[compress]{cite}
\usepackage{amssymb}   
\usepackage{bm} 
\usepackage{xcolor}
\usepackage{hyperref}
\hypersetup{
	colorlinks   = true, 
	urlcolor     = blue, 
	linkcolor    = blue, 
	citecolor   = red 
}

 \def\g{\sqrt{-g}\,}

\def\p{\partial}

\title{A novel probe of Einstein-Hilbert action: Dynamic upgradation of metric parameters}

\author{Krishnakanta Bhattacharya\footnote{krishnakanta@iucaa.in} \\
IUCAA, Post Bag 4, Ganeshkhind,
 Pune - 411 007, India.}

\date{\today}

\begin{document}

\maketitle

\begin{abstract}
The Einstein-Hilbert (EH) action is peculiar in many ways. Some of the Peculiar features have already been highlighted in literature. In the present article, we have discussed some peculiar features of EH action which has not been discussed earlier. It is well-known that there are several ways of decomposing the EH action into the bulk and the surface part with different underlying motivations. We provide a review on all of these decompositions. Then, we attempt to study the static coordinate as a limiting case of a time-dependent coordinate via dynamic upgradation of the constant metric parameters. Firstly, we study the consequences when the constant parameters, present in a static and spherically symmetric (SSS) metric, are promoted to the time dependent variables, which allows us to incorporate the time-dependence in the static coordinate. We find that, in every sets of decomposition, the expression of the bulk term remains invariant, whereas the surface term changes by a total derivative term. Finally, when we obliterate the time dependence of the metric parameters, we find that the expression of the Ricci-scalar (or the EH action) does not go back to its original value. Instead, we find that the curvature becomes singular on the horizon, which implies a topological change from the original spacetime.
\end{abstract}
\section{Introduction}

All the fundamental theories are expected to be obtained via a well-defined action principle. In general relativity, the Einstein-Hilbert (EH) action is widely accepted as one which provides the dynamics of the gravitational system. However, there have been debates for long regarding the well-posedness of the action principle while obtaining the Einstein's field equation from the EH action. In addition, there are several peculiar features of EH action which makes it distinctively different from the actions of other fundamental theories. In this case, the action contains the second order derivative of the metric. Therefore, one has to fix both the metric as well as its first order derivative on the boundary (\textit{i.e.} one has to impose both Dirichlet as well as Neumann boundary condition simultaneously), which creates the action principle to be devoid of being a well-posed formulation. This issue can be resolved by adding a suitable boundary term \cite{Charap:1982kn} along with the EH action. The most popular boundary term in literature is the Gibbons-Hawking-York (GHY) boundary term \cite{York:1972sj,Gibbons:1976ue,York:1986lje} (there are several other popular boundary term, for ref see \cite{Charap:1982kn}). In addition, the peculiar structure of the EH action helps to get rid of such problem. The second order derivative terms of the EH action can be expressed as a total derivative as a whole and, therefore, does not contribute in the dynamics. Thus, the EH action can be decomposed into bulk part as well as the surface part, where the bulk part contains the first order derivative of the metric and the surface term contains the second order derivative of the metric. Most importantly, it has been found by Padmanabhan and collaboration \cite{Padmanabhan:2002jr,Padmanabhan:2002xm,Padmanabhan:2003gd,Padmanabhan:2004fq,Padmanabhan:2006fn,Mukhopadhyay:2006vu} that the bulk part and the surface part of the EH action are not independent of each other. They are related by the ``holographic relation'', which makes gravity intrinsically holographic as this relation suggests that the surface degrees of freedom participates in the dynamics of the system.  Apart from this decomposition, there are several other decomposition in literature such as the ADM decomposition and the decomposition in terms of $G^0_0$ and $R^0_0$. Each of these decompositions of EH action has its own significance, which has been discussed briefly in this article.

 The scope of the present paper is to discuss about the EH action, its decompositions and its peculiar features which has not been discussed earlier. The effect of coordinate transformation in the expression of EH action and in bulk/ surface part of different decompositions are quite known (for example see \cite{Boehmer:2021aji}). In this paper, we formulate a novel approach to probe the EH action and its decomposed parts.  The idea is that a static metric can be considered as a limiting case of a time-dependent one. For example, there are several constant parameters which appears in the spacetime metric. We can promote (some/ all of) those constant parameters as time-dependent variables and think the static metric as a limiting case of this time-dependent metric, where the static metric can be obtained in the limit where the promoted time-dependent parameters corresponds to its earlier constant value. Of course, these two metrics (one with constant metric parameters and other with the time-dependent parameter(s)) will correspond to different spacetime geometry and different expressions of the dynamical equations. However, we can check how the expression of the EH action (and the bulk or the surface term of different decompositions) changes (change) due to such change in the parameters. More importantly, it will be interesting to see whether the spacetime geometry/ EH action boils down to its static expression in the limit where the promoted parameters correspond to their constant value (\textit{i.e.} the question we ask is that if we make a reversible change in the metric (via dynamic upgradation), whether the spacetime geometry changes reversibly).

  Here, it will be shown that in static and spherically symmetric (SSS) metric, if one incorporates the time-dependence by promoting the underlying constant parameters as time-dependent variables, the expression of the EH action changes by a total time-derivative term. When we make a deeper investigation, we find that the bulk part of each decomposition remains  unchanged and only the surface part of each decomposition changes by that total derivative term. This characteristics of the EH action (\textit{i.e} the bulk part of the Lagrangian in different decomposition being invariant under the dynamic upgradation of constant metric parameters) can be attributed to spherically symmetric metric. On the other hand, as we show later, in Painleve coordinate, both the bulk as well as the surface part changes under such dynamic upgradation of parameters and the invariance of the bulk Lagrangian is no longer preserved. When the time dependence of the constant parameters are removed (by introducing a smallness parameter $\epsilon$ and taking the limit $\epsilon\rightarrow 0$, in which the time dependent parameter will again boil down to its constant value), one expects that the expression of the EH action will reduce to its original expression. However, we have found that the curvature (or the EH Lagrangian) becomes singular with infinite discontinuity near the horizon. This implies a permanent and non-reversible shift of the spacetime geometry due to dynamic upgradation.
 
 The paper is organized as follows. In the following section (\textit{i.e.} section \ref{mathbg}) we provide all the decompositions of the EH action and mention its significance briefly. In section \ref{DUP} we describe the consequence (in EH action) when the constant parameters in the metric is promoted to a time-dependent variables in SSS metric. In section \ref{OBLIT}, we obliterate the time dependence and check whether the EH action reduces to its original value. Finally, in section \ref{COPMA} we provide a comparative discussion on coordinate transformation and dynamic upgradation. The conclusion of the analysis is provided in section \ref{CONCLU}.

\section{Mathematical background: Avatars of Einstein-Hilbert action} \label{mathbg}
The starting point of the present analysis is the Einstein-Hilbert (EH) action, which is given as 
\begin{align}
A_{EH}=\int_{\nu} \g L d^4x=\frac{1}{16\pi}\int_{\nu} \g R d^4x~. \label{EHACT}
\end{align}
Not only the above action describes the dynamics in the curved spacetime, several other features of gravity are also encoded in the above action. To explore those features, the above form of the action is expressed in several ways. In this section, we briefly discuss about those several forms of the action, their implications, and, make a comparative analysis of those different forms of the action. 

The Einstein-Hilbert action \eqref{EHACT} contains the first-derivative as well as the second-derivative of the metric tensor $g_{ab}$. As a result, the principle of extremal action becomes ill-defined when it is followed from the action \eqref{EHACT}. This is because, in that case, one has to fix both the metric and its first-derivative on the boundary. To resolve this issue, there are two different routes. The first one is the addition of a suitable surface term with the action \eqref{EHACT} so that its variation cancels the aforementioned first-derivative, which were required to be fixed on the boundary. The most popular surface term in the literature is the Gibbons-Hawking-York (GHY) boundary term \cite{York:1972sj,Gibbons:1976ue,York:1986lje}. On a surface, which has normal $u^a$, the GHY term is defined as 
\begin{align}
A_{GHY}=\frac{\bar \epsilon}{8\pi}\oint_{\p\nu}\sqrt{h_{\perp}}Kd^3y~, \label{GHY}
\end{align}
where, $\bar \epsilon=u^iu_i=\pm1$ ($+1$ for time-like and $-1$ for space-like surface respectively), $h_{\perp}$ is the determinant of the induced metric on the surface and $K=-\nabla_iu^i$ is the trace of the extrinsic curvature on the surface. The above surface term, can be equivalently written as a bulk term as $A_{GHY}=(1/8\pi)\int_{\nu}\g\nabla_i(Ku^i)d^4x$ which, by using Gauss' theorem, boils down to Eq. \eqref{GHY}. For a space-like hypersurface, we denote the normal as $u^i\equiv n^i$, $h_{\perp}\equiv h$ and $K=-\nabla_in^i$ (this notational convention we shall follow during the discussion of ADM decomposition). Usually, people add the GHY term \eqref{GHY} with the action \eqref{EHACT} and show that the problematic part (\textit{i.e.} the first order derivative of the metric, which were required to be fixed on the boundary) gets cancel with the variation of the GHY term. However, it has be systematically proved by Padmanabhan \cite{Padmanabhan:2014lwa} that if one starts from the Einstein-Hilbert action \eqref{EHACT}, one needs to add the GHY term \eqref{GHY} in order to make the principle of extremal action well-defined. More importantly, it can also be shown that all components of the metric are not required to be fixed on the boundary. For instance, for $t=$constant boundary, only the spatial part of the metric (\textit{i.e.} $g_{\alpha\beta}$, where the Greek indices $\alpha$, $\beta$ \textit{etc.} indicates the spatial components) are required to be fixed, whereas the components $g_{00}$ and $g_{0\alpha}$ can be left free.

 Although, the GHY boundary term is a popular boundary term, it cannot be applied for the null surfaces as the construction of the GHY term requires non-degenerate induced metric on the boundary and unit normals. Thus, the definition of a suitable boundary term for a null surface is more non-trivial, which has been resolved from the works of Padmanabhan and collaboration, who defined a proper counter term on the null surface as $2\sqrt{q}(\Theta+\kappa)$ \cite{Parattu:2015gga}, where $\Theta$ is the expansion parameter of the null-geodesics on the null surface, $q$ is the determinant of the two-metric on the null surface and $\kappa$ is the surface gravity. This has been further generalized, in order to make it applicable for both null and non-null surfaces \cite{Parattu:2016trq,Chakraborty:2016yna,Jubb:2016qzt,Chakraborty:2018dvi}.

Above, we have discussed the procedure of adding a proper counter term which cancels the problematic term on the boundary while taking variation of the EH action. In literature, there is another procedure, which we discuss in more detail in the following. There are several prescriptions in literature to decompose the EH action into the bulk and the surface part. In the following, we shall discuss those decompositions one-by-one along with their significances.
 
 \subsection{Decomposition I}

  The above action \eqref{EHACT} can be decomposed into the two parts: one is the quadratic part, which contains the first derivative of the metric tensor $g_{ab}$ and another one is the surface part, which is a total-derivative term and contains the second derivative of $g_{ab}$ (extensively discussed in the chapter 6 of \cite{gravitation}). It can be shown that this decomposition helps us to obtain the Einstein's equation in a well-defined manner in two different ways. 1. One can drop the surface part of the action (as it is a total derivative term and, hence, presence of this part does not alter the dynamics of the system) and the equation of motion is obtained only from the bulk part. 2. One can define proper conjugate variables and can obtain Einstein's equation from the whole EH action by fixing the conjugate momenta on the boundary. The decomposition of the Einstein-Hilbert action, in terms of the quadratic part and the surface part, is described as follows.
\begin{align}
16\pi\g L=\g R=\g L_{quad}+L_{sur}~. \label{DECOMP1}
\end{align}
The quadratic term, containing the first-order derivative of the metric tensor, is given by the following expression.
\begin{align}
L_{quad}=g^{ab} \Big(\Gamma^i_{ja}\Gamma^j_{ib}-\Gamma^i_{ab}\Gamma^j_{ij}\Big)=2Q_a^{\ bcd}\Gamma^a_{dk}\Gamma^k_{bc}~, \label{LQUAD1}
\end{align}
where 
\begin{align}
Q_a^{\ bcd}=\frac{\p R}{\p R^a_{\ bcd}}=\frac{1}{2} (\delta^c_ag^{bd}-\delta^d_ag^{bc})~,
\end{align}
which has the same symmetric/anti-symmetric property as of the Riemann tensor due to the exchange of indices.
%
 
 The surface part of the Einstein-Hilbert action, as mentioned in Eq. \eqref{DECOMP1} is given as
 \begin{align}
 L_{sur}\equiv\p_c(\g V^c), \ \ \ \textrm{where}\ \ \ V^c\equiv(g^{ik}\Gamma^c_{ik}-g^{ck}\Gamma^m_{km})=-\frac{1}{g}\p_b(gg^{bc})~, \label{LSUR}
 \end{align}
 where $g$ is the determinant of the metric tensor $g_{ab}$. 
 
 Since all the second order derivatives of the metric is within $L_{sur}$, which is a total derivative term, one can obtain the Einstein's equation from the bulk part only (see chapter 6 of \cite{gravitation}). Thus, this peculiar structure of EH Lagrangian helps us to formulate a well-posed action principle. Remember, that the decomposition of the Einstein-Hilbert action in terms of the quadratic part (also known as the bulk Lagrangian) and the surface part is not a covariant one and the quadratic and the surface Lagrangian are not covariant scalars. Interestingly, the quadratic and the surface Lagrangian are not independent of each other. Instead, those are connected to each other by the following relation, which is known as the \textit{holographic relation} in the work of Padmanabhan et. al. \cite{Padmanabhan:2002jr,Padmanabhan:2002xm,Padmanabhan:2003gd,Padmanabhan:2004fq,Padmanabhan:2006fn,Mukhopadhyay:2006vu,Kolekar:2010dm}, which is given as
\begin{align}
L_{sur}=-\p_c\Big[g_{ab}\frac{\p(\g L_{quad})}{\p(\p_c g_{ab})}\Big]~. \label{HOLOGRAPHY}
\end{align}
The above holographic relation also implies that the gravity is intrinsically Holographic as the surface degrees of freedom are related to the dynamical degrees of freedom . 

The above connection of the quadratic and the surface Lagrangian helps us to interpret the Einstein-Hilbert Lagrangian as the Lagrangian of the momentum space by the following analogy with the classical mechanics. Consider the Lagrangians $L_1\equiv L_1(q^A,\p q^A)$ and $L_2\equiv L_2(q^A,\p q^A, \p^2q^A)$. If $L_1$ and $L_2$ are connected by the relation $L_2(q^A,\p q^A, \p^2q^A)=L_1(q^A,\p q^A)-\p_i(q^AP^i_A)$ (where, the canonical momenta $P^i_A$ is defined as $P^i_A=\p L_1/\p (\p_iq^A)$), it can be shown that $L_1$ and $L_2$ yields the same equation of motion. However, when the Lagrangian is $L_1$, one has to fix $q^A$ on the boundary whereas, when the Lagrangian is $L_2$, one has to fix $P^i_A$ on the boundary. 
It can be shown that the Einstein-Hilbert Lagrangian $L$ corresponds to $L_2(q^A,\p q^A, \p^2q^A)$ and the quadratic Lagrangian $\g L_{quad}$ corresponds to $L_1(q^A,\p q^A)$ of the above discussion. Also, $\p(\g L_{quad})/\p(\p_cg_{ab})=\g M^{cab}$ corresponds to the canonical momentum of $g_{ab}$. Similarly $-\g M^a_{bc}$ corresponds the momentum of $g^{bc}$ (for detail discussion, see \cite{Parattu:2013gwa}). Importantly, one can either obtain equation of motion from the $L_{quad}$ only by fixing $g^{ab}$ at the boundary, or one can take the whole Einstein-Hilbert Lagrangian and obtain equation of motion by fixing $\g M^c_{ab}$ at the boundary. Thus, $g^{ab}$ and $\g M^c_{ab}$ act as a pair of holographically conjugate variables (HCV's) \cite{Parattu:2013gwa}. Apart from $g^{ab}$ and $\g M^c_{ab}$, another pair of HCV's can be introduced as $f^{ab}$ and $N^c_{ab}$, where $f^{ab}=\sqrt{-g}g^{ab}$ and $N^c_{ab}$ is the conjugate momentum (see \cite{Parattu:2013gwa}, also see \cite{Chakraborty:2014joa} for the generalization to Lanczos-Lovelock gravity). These definitions also interprets the EH action as the action of momentum space. In addition, these definition of $f^{ab}$ and $N^c_{ab}$ also help us to obtain the field equations in a similar way as of the Hamilton's equation of classical mechanics \textit{i.e.} $\p_cf^{ab}=\p \mathcal{H}/\p N^c_{ab}$ and $p_cN^c_{ab}=-\p\mathcal{H}/\p f^{ab}$, where $\mathcal{H}$ represents the suitably defined Hamiltonian \cite{Parattu:2013gwa,Chakraborty:2014joa}.

Although this method of decomposition helps us to define a well-posed action principle and to illuminate the holographic nature of gravity, this method is not a covariant one. In the following, we shall discuss about other decompositions which are covariant but also are foliation dependent.

\subsection{Decomposition II}
A proper canonical formalism of general relativity is provided by the ADM (named after Arnowitt, Deser and Misner) formalism \cite{Arnowitt:1962hi}, which depends on the  foliation of the spacetime (also see \cite{gravitation, Chakraborty:2016yna}). In this case, it is considered that the spacetime is foliated by the family of space-like surfaces with time-like normal $n_a$ (which is normalized \textit{i.e.} $n^an_a=-1$) and induced metric on the spacelike surface is defined as $h_{ab}=g_{ab}+n_an_b$. Then, the Einstein-Hilbert action can be written in terms of the ADM Lagrangian as 
\begin{align}
16\pi \g L=\g R=\g L_{ADM}+L_{sur}^{(ADM)}~,
\end{align}
where, the ADM Lagrangian $L_{ADM}$ is given by
\begin{align}
L_{ADM}=^{(3)}R+K_{ab}K^{ab}-K^2~, \label{LADM}
\end{align}
and the surface term $L_{sur}^{(ADM)}$ is given as 
\begin{align}
L_{sur}^{(ADM)}=-2\g\nabla_i(Kn^i+a^i)~. \label{LSURADM}
\end{align}
Here $^{(3)}R$ is the three-Ricci scalar corresponding to the metric $h_{\alpha\beta}$ (where $\alpha$, $\beta$ denote the spatial indices); $K_{ab}$ is the extrinsic curvature tensor of the spacelike surface, which is given by $K_{ab}=-h^m_ah^n_b\nabla_mn_n$; $K$ is the trace of $K_{ab}$ and $a^i=n^a\nabla_an^i$~.

Again, in this case, one can obtain Einstein's equation either from the bulk ADM action (where $h^{\alpha\beta}$ are required to be fixed on the boundary) or from the total action (by fixing the corresponding momenta of $h^{\alpha\beta}$ on the boundary). This method is also important because this provides the Hamiltonian formalism of general relativity, which plays an important role in canonical quantum gravity and numerical relativity.

As we have discussed, one can obtain the Einstein's equation from the bulk part of these decompositions ($L_{quad}$ and $L_{ADM}$). Whereas both $L_{sur}$ and $L_{sur}^{(ADM)}$ are the surface terms,Presence of which does not alter the dynamics of the system. However, presence of these surface terms in the action (in both the decomposition) interprets the EH action as the action of the momentum space as one has to fix the corresponding momentum in the boundary. However, note that $L_{sur}$ and $L_{sur}^{(ADM)}$  are not equal and, therefore, $L_{quad}\neq L_{ADM}$. The spacetime foliated by $t=$constant hypersurfaces are properly described by the following metric:
\begin{align}
ds^2=-N^2dt^2+h_{\alpha\beta}(dx^{\alpha}+N^{\alpha}dt)(dx^{\beta}+N^{\beta}dt)~. \label{ADMMETRIC}
\end{align}
For this metric, one can obtain 
\begin{align}
L_{sur}-L_{sur}^{(ADM)}=\g(L_{ADM}-L_{quad})=\p_0\Big(\frac{\sqrt{h}}{N}\Big)(\p_{\alpha}N^{\alpha})-\p_{\alpha}\Big(\frac{\sqrt{h}}{N}\Big)(\p_0N^{\alpha})
\nonumber
\\
-\p_{\alpha}\Big[\frac{N}{\sqrt{h}}\p_{\beta}(hh^{\alpha\beta})\Big]+\p_{\alpha}\Big(\frac{\sqrt{h}}{N}\Big)\Big[N^{\beta}\p_{\beta}N^{\alpha}-N^{\alpha}\p_{\beta}N^{\beta}\Big]
\nonumber
\\
+\frac{\sqrt{h}}{N}\Big[(\p_{\alpha}N^{\beta})(\p_{\beta}N^{\alpha})-(\p_{\alpha}N^{\alpha})(\p_{\beta}N^{\beta})\Big]~.
\end{align}
Also note that the GHY term is not the same as either of $L_{sur}$ or $L_{sur}^{(ADM)}$. The GHY Lagrangian, on a space-like surface can be written as $16\pi L_{GHY}=2\nabla_i(Kn^i)$ (where $A_{GHY}=\int_{\nu}\g L_{GHY}d^4x$) and for the general metric \eqref{ADMMETRIC}, it can be shown that it is not the same as the $L_{sur}$ or $L_{sur}^{(ADM)}$.
\subsection{Decomposition III}
There is another way of decomposing the Einstein-Hilbert Lagrangian with the help of the normal $n^a$, which is given by \cite{gravitation,Padmanabhan:2004fq}
\begin{align}
16\pi \g L=\g R=2\g(G_{ab}-R_{ab})n^an^b \label{DECOMP3}
\end{align}
From the Gauss-Codazzi relations we obtain
\begin{align}
2G_{ab}n^an^b=^{(3)}R-K_{ab}K^{ab}+K^2~.\label{GABNANB}
\end{align}
When Einstein's equation holds, $2G_{ab}n^an^b=16\pi\rho$ can be identified as the numerical value of ADM Hamiltonian density. Therefore, it can be considered as the energy density.
 Now, $R_{ab}n^an^b$ is given by
\begin{align}
R_{ab}n^an^b=\nabla_i(Kn^i+a^i)-K_{ab}K^{ab}+K^2~.
\end{align}
In a static spacetime, above decomposition of the Einstein-Hilbert action (as mentioned in \eqref{DECOMP3}), has a particular significance from thermodynamic viewpoint. It interprets the Einstein-Hilbert action as the free energy of the spacetime \cite{Padmanabhan:2004fq}. The near-horizon geometry of a static spacetime can be described by the metric \cite{Medved:2004ih}
\begin{align}
d\bar s^2=-N^2dt^2+dl^2+\sigma_{AB}dx^Adx^B; \ \ \ N=\kappa l+\mathcal{O}(l^3); \ \ \ \sigma_{AB}=\mu_{AB}(x^A)+\mathcal{O}(l^2)~, \label{GENSTAT}
\end{align}
where, the location of the horizon is given by $l=0$~; $A$, $B$ \textit{etc.} are the indices of the two surface and $\kappa$ is the surface gravity. For the above metric \eqref{GENSTAT} (in general, it is true for any arbitrary static metric, expressed in coordinate system with no shift, \textit{i.e.} for $N^{\alpha}=0$) one can obtain that all the components of $K_{ab}$ vanishes. Therefore, one finds 
\begin{align}
R_{ab}n^an^b=-R^0_0=\nabla_ia^i=\frac{1}{\g}\p_{\alpha}(\g a^{\alpha})~,
\end{align}
which implies that $R^0_0$ is a total derivative term. In addition, the fact, that $R^0_0$ is a total derivative term in a static spacetime, can also be proved generally in a metric independent way. For static spacetime, one can consider the presence of time-like Killing vector $\xi^a=\{1,0,0,0\}$. Therefore, one can obtain \cite{Padmanabhan:2004fq}
\begin{align}
R^a_i\xi^i=R^a_0=\nabla_b(\nabla^a\xi^b)=\frac{1}{\g}\p_b(\nabla^a\xi^b)~.
\end{align}
It yields that all the components of $R^a_0$ are total-derivative term in static spacetime and, the expression of $R^0_0$ is given as  $R^0_0=\p_{\alpha}(\g g^{i0}\Gamma^{\alpha}_{i0})/\g$. Therefore, for any static metric with no shift (such as \eqref{GENSTAT}), $a^{\alpha}=-g^{0i}\Gamma^{\alpha}_{0i}$). It is also noteworthy that for any static spacetime, expressed in coordinate system with no shift, one obtains
\begin{align}
L_{sur}^{(ADM)}=-2\g R_{ab}n^an^b=2\g R^0_0=-2\g\nabla_ia^i=2\p_{\alpha}(\g g^{i0}\Gamma^{\alpha}_{i0})~.
\end{align}
In that case (\textit{i.e.} in a static spacetime with no shift), one also finds $L_{ADM}=2G_{ab}n^an^b$. However, even that case, $L_{sur}$ does not coincide with either $L_{sur}^{(ADM)}$ or with $-2\g R_{ab}n^an^b$. 

Interestingly, in spacetime \eqref{GENSTAT}, $L_{sur}$, $L_{sur}^{(ADM)}$ and $-2\g R_{ab}n^an^b$-- all of the surface terms have the thermodynamic interpretation as the entropy on the horizon surface ($l=0$). Also, in any spacetime, $2G_{ab}n^an^b$ can be identified as the energy density, which we have discussed earlier. Therefore, the decomposition of Einstein-Hilbert action, as stated in equation \eqref{DECOMP3}, interprets the Einstein-Hilbert action as the free energy of the spacetime \cite{Padmanabhan:2004fq,Kolekar:2010dm} (\textit{i.e.} $A_{EH}\equiv\beta E-S$), where the integration containing $G_{ab}n^an^b$ can be interpreted as $\beta E$ (with $\beta$ being the periodicity of Euclidean time, which is identified as inverse temperature) and the integration containing $R_{ab}n^an^b$ can be interpreted as the entropy. Furthermore, when $A_{sur}=\int L_{sur}d^4x$ (where $L_{sur}$ has been defined while discussing ``Decomposition I'') when integrated over the horizon of a static metric, provides $\tau TS$, where $\tau$ is the range of the time-integration, $T$ is the Hawking temperature and $S$ is the Bekenstein-Hawking entropy \cite{Mukhopadhyay:2006vu,Padmanabhan:2012qz}. This leads to the further definition of ``surface Hamiltonian'' as $H_{sur}=-\p A_{sur}/\p \tau=TS$ \cite{Padmanabhan:2012qz}~. In addition, the term $pdq$ of ``Decomposition I'' (\textit{i.e.} $N^c_{ab}\delta f^{ab}$ or $\sqrt{-g}M^c_{ab}\delta g^{ab}$) integrated over the transverse surface (for the general static metric \eqref{GENSTAT}) has the thermodynamic interpretation of $T\delta S$, while $qdp$ (\textit{i.e.} $f^{ab}\delta N^c_{ab}$ or $\delta(\sqrt{-g}M^c_{ab}) g^{ab}$) has the thermodynamic interpretation of $S\delta T$ \cite{Parattu:2013gwa} (also see \cite{Chakraborty:2014joa} for the generalization to Lanczos-Lovelock gravity). Also, since the total EH action can be expressed as the free energy of the spacetime, the action principal can be interpreted as the thermodynamical extremum principal (for example, see ref. \cite{Padmanabhan:2004fq,Parattu:2013gwa}).


Note that both ``Decomposition II'' and ``Decomposition III'', which we have discussed above, crucially depends upon the timelike (or spacelike) nature of the boundary surface and, these are not adapted for the null-surfaces. For Hamiltonian formulation on the null-surfaces, see \cite{Torre:1985rw,Goldberg:1995gb}. Earlier, we have commented that, in static spacetime, the surface parts of each decomposition (including ``Decomposition II'' and ``Decomposition III'') have the thermodynamic interpretation as the entropy on the horizon surface (and the bulk as thermodynamic energy). Now, the horizon is a null surface in general. Nevertheless, the thermodynamic interpretation of the bulk and the surface terms remain valid (including in the case of ``Decomposition II'' and ``Decomposition III''). This is because, the black hole horizon in static spacetime is also the Killing horizon (as per Hawking's rigidity theorem). The normal of a Killing horizon, \textit{i.e.} the Killing vector, is usually non-null everywhere except on the horizon. Thus, ``Decomposition II'' and ``Decomposition III'' is valid for the Killing normals and can have thermodynamic interpretations on the horizon.

Before moving on to the next part, we mention that the similar decompositions and their connection with each other of the Einstein-Hilbert action has been presented earlier in literature (for example see \cite{Parattu:2013gwa, Chakraborty:2016yna} \textit{etc.}). So far, we have mentioned how the Einstein-Hilbert action is decomposed in several ways and also have mentioned about the implications of each form of the action. In the following section, we shall investigate what happens when the constant parameters in a static spacetime is promoted to time-dependent variables.


\section{Dynamic upgradation of metric parameters} \label{DUP}
For static metric, all the time-derivatives vanish and all the expressions (Einstein-Hilbert action and its several decomposed forms) appear in terms of the spatial derivatives. Therefore, one can think that these static metric can be viewed as a limiting case of the time-dependent metric. From now onward, we want to probe how the expression of the EH action of a time-dependent metric differs from its static counterpart. Also, it will help us to find whether it is a right way to think the static metric as a limiting case of the time-dependent one. To analyse these, the method we adopt is the following, which we call as the dynamic upgradation of metric parameters. In general, the spacetime metric contains several parameters. It is therefore, interesting to ask that what will happen if we promote those constant parameters to the time-dependent variables. This will allow us to introduce the time-dependence in the static spacetime. In addition, it will allow us to check the consequence in the limit where the promoted parameters are reduced to the constants again (\textit{i.e.} the metric again becomes static). Note, here we are interested in the level of EH action (or the reduced EH action) and we do not bother about the underlying dynamics (or EH equation). In other words, the analysis which we follow here is off-shell.

Let us start from a particular example. A Schwartzschild metric is described by the metric $\bar{g}_{ab}\equiv\bar{g}_{ab}(r,M)$ (from now onward, a bar overhead will imply a static metric). If we promote $M\longrightarrow M(t)$, the metric will be given as 
\begin{align}
ds^2=g_{ab}dx^adx^b=-\Big(1-\frac{2M(t)}{r}\Big)dt^2+\frac{dr^2}{\Big(1-\frac{2M(t)}{r}\Big)}+r^2d\Omega^2~. \label{TDEPSCH}
\end{align}
The Ricci-scalar corresponding to $\bar{g}_{ab}(r,M)$ is $\bar{R}=0$. However, the same for the metric \eqref{TDEPSCH} is given by 
\begin{align}
\g R=\frac{d}{d t}\Big[\frac{2\dot{M}r\sin\theta}{(1-\frac{2M}{r})^2}\Big]~.  \label{TRSCH}
\end{align}
Clearly, the Schwartzschild metric and the metric mentioned in \eqref{TDEPSCH} describe different spacetimes with different geometries. Also, it must be mentioned that the source of the metric \eqref{TDEPSCH} is rather unphysical as the energy-momentum tensor ($T_{ab}$) corresponding to the metric has only non-vanishing components $T_{\theta\theta}$, $T_{\phi\phi}$ and $T_{tr}$. Since $T_{tt}$ and $T_{rr}$ vanishes, it corresponds to a system with zero energy and non-zero transverse pressure, which implies that the source is unphysical. However, since we are not concerned about solving Einstein's equation, we do not bother about it. On the other hand, another example of dynamic upgradation of mass, which lead to the physical solution, is the Vaidya metric \cite{Vaidya:1999zza}. Vaidya metric can be obtained in the following way. From Schwarzschild metric, one can make a coordinate transformation by replacing the (Schwarzschild) time coordinate with the advanced/ retarded null coordinate. Thereafter, the constant mass parameter is promoted to the function of the advanced/ retarded null coordinate. Physically, Vaidya metric represents spherically symmetric body radiating or absorbing null dusts.

 Above conclusion (\textit{i.e.} the reduced EH action becoming a total-derivative) can also be drawn for Reissner-N$\ddot{\textrm{o}}$rdstrom (RN) spacetime. If we take the RN metric $\bar{g}_{ab}(r, M, Q)$ and promote the $M$ and $Q$ as $M\longrightarrow M(t)$ and $Q\longrightarrow Q(t)$, the metric will be given as 
\begin{align}
ds^2=g_{ab}dx^adx^b=-\Big(1-\frac{2M(t)}{r}+\frac{Q(t)^2}{r^2}\Big)dt^2+\frac{dr^2}{\Big(1-\frac{2M(t)}{r}+\frac{Q(t)^2}{r^2}\Big)}+r^2d\Omega^2~. \label{TDEPRN}
\end{align}
 The Ricci-scalar of RN metric $\bar{g}_{ab}(r, M, Q)$ is $\bar{R}=0$, whereas for metric \eqref{TDEPRN} it is given by 
\begin{align}
\g R=\frac{d}{d t}\Big[\frac{2(\dot{M}r-Q\dot{Q})\sin\theta}{(1-\frac{2M}{r}+\frac{Q(t)^2}{r^2})^2}\Big]~.
\end{align}
So far, all the results are obtained for particular spacetimes such as Schwartzschild spacetime, RN spacetime. It can generally be proved that for a general static and spherically symmetric spacetime, one can obtain a similar conclusion.

A general static and spherically symmetric (SSS) metric with constant parameters ($\kappa$'s and $\lambda$'s) is given as 
\begin{align}
d\bar{s}^2= -f(r, \kappa)dt^2+\frac{dr^2}{g(r,\lambda)}+r^2d\Omega^2~. \label{SSSFG}
\end{align}
The Ricci-scalar corresponding to the metric \eqref{SSSFG} is given as 
\begin{align}
\bar{R}=\frac{1}{2r^2f^2}\Big[r^2gf'^2-4f^2(-1+g+rg')-rf(rf'g'+2g(2f'+rf''))\Big]~.\label{RICCISSSFG}
\end{align}
Now, if we promote the constant parameters to time-dependent variables, one can obtain the metric as
\begin{align}
ds^2= -f(r, \kappa(t))dt^2+\frac{dr^2}{g(r,\lambda(t))}+r^2d\Omega^2~. \label{SSTFG}
\end{align}
The expression of Ricci-scalar corresponding to the metric \eqref{SSTFG} is provided as
\begin{align}
\g R=\g\bar{R}\Big|_{\kappa,\lambda\longrightarrow\kappa(t),\lambda(t)}-\frac{d}{dt}\Big[\frac{r^2\sin\theta\dot{\lambda}_i(\p g/\p\lambda_i)}{\sqrt{f}g^{3/2}}\Big]~, \label{DIFFEHLAG}
\end{align}
 where $\bar{R}|_{\kappa,\lambda\longrightarrow\kappa(t),\lambda(t)}$ implies the expression of $\bar{R}$ (as provided in Eq. \eqref{RICCISSSFG}) with $\kappa$'s and $\lambda$'s been promoted to $\kappa(t)$'s and $\lambda(t)$'s respectively. Therefore, one can conclude that the expression of Einstein-Hilbert Lagrangian differ by a total time-derivative term when the constant parameters of a static \& spherically symmetric metric is promoted to time-dependent variables.

  Interestingly, as we discuss it in the later part of the section, the total derivative term (\textit{i.e.} the last term of \eqref{DIFFEHLAG}) can be identified as $-2\g\nabla_i(Kn^i)$ for the metric \eqref{SSTFG}. Therefore, one obtains
\begin{align}
\frac{1}{16\pi} \g R=\frac{1}{16\pi}\g\bar{R}\Big|_{\kappa,\lambda\longrightarrow\kappa(t),\lambda(t)}-\frac{1}{8\pi}\g\nabla_i(Kn^i)~. \label{DIFFEHLAG2}
\end{align} 
The last term of \eqref{DIFFEHLAG2} can be identified as the negative of GHY Lagrangian on a spacelike hypersurface.
 Also, note that the difference in the Lagrangian, as it has been mentioned in \eqref{DIFFEHLAG}, appears due to the fact that the constant parameters ($\lambda$'s) in $g(r,\lambda)$ are promoted to time-dependent variables and it does not depend on whether the constant parameters in $f(r, \kappa)$ are promoted to time dependent variables. Therefore, in metric \eqref{SSSFG}, if we promote $\kappa$'s as $\kappa\longrightarrow\kappa(t)$ and keep $\lambda$'s unchanged, the extra total-time-derivative term in \eqref{DIFFEHLAG} will not appear and the Einstein Hilbert Lagrangian will be exactly same even after promoting $\kappa$'s to time-dependent variables.
 
Earlier, we have discussed how Einstein-Hilbert action can be decomposed in several forms. We shall now investigate how the decomposed components of Einstein-Hilbert action changes due to the dynamic upgradation of constant parameters. As we have discussed earlier, the quadratic part of Einstein-Hilbert Lagrangian plays the pivotal role while obtaining the Einstein's equation from the action principle. The quadratic part of the Lagrangian corresponding to the metric \eqref{SSSFG} is obtained as 
\begin{align}
\bar{L}_{quad}=\frac{2g(f+rf')}{r^2f}~.
\end{align}
After promoting constant parameters to time-dependent variables, one can obtain the same quantity corresponding to the metric as 
\begin{align}
L_{quad}=\bar{L}_{quad}\Big|_{\kappa,\lambda\longrightarrow\kappa(t),\lambda(t)}~,
\end{align}
which means that the quadratic part of the Einstein-Hilbert Lagrangian remains invariant even after upgradation of parameters to time-dependent variables in a SSS coordinate. Since the quadratic part remains unchanged, and the Einstein-Hilbert Lagrangian differ by a total time-derivative term (as mentioned in \eqref{DIFFEHLAG}), the surface part of the Lagrangian will also differ by the same total derivative term \textit{i.e.,}
\begin{align}
L_{sur}=\bar{L}_{sur}\Big|_{\kappa,\lambda\longrightarrow\kappa(t),\lambda(t)}-\frac{d}{dt}\Big[\frac{r^2\sin\theta\dot{\lambda}_i(\p g/\p\lambda_i)}{\sqrt{f}g^{3/2}}\Big]~. \label{DIFFLSUR}
\end{align}
Again, if we upgrade only $\kappa$'s to time-dependent variables and leave $\lambda$'s unchanged, the surface term will also remain unchanged due to $\kappa\longrightarrow\kappa(t)$.

Let us investigate now about the ADM decomposition of the Einstein-Hilbert action. If we compute the ADM Lagrangian, which is defined in \eqref{LADM}, for the SSS metric \eqref{SSSFG}, one can find, the expression of the ADM Lagrangian is given as $L_{ADM}=^{(3)}R$ (due to the fact that all components of $K_{ab}=0$.) Since the expression of $^{(3)}R$ does not contain any time-derivative, its expression does not change when the parameters are promoted to time-dependent variables. On the contrary, when the parameters are promoted, the only non vanishing component of $K_{ab}$ (corresponding to the metric \eqref{SSTFG}) is $K_{rr}$~. As a result,  $K_{ab}K^{ab}-K^2=(g^{rr})^2(K_{rr}K_{rr}-K_{rr}K_{rr})=0$~. Therefore, we obtain that the expression of the ADM Lagrangian is also unchanged when the parameters are promoted to time-dependent variables in a general SSS spacetime, \textit{i.e.,}
\begin{align}
L_{ADM}=\bar{L}_{ADM}\Big|_{\kappa,\lambda\longrightarrow\kappa(t),\lambda(t)}~.
\end{align}
Following the same argument, as been provided for $L_{sur}$, we obtain that the surface part of ADM decomposition ($L_{sur}^{(ADM)}$) changes in a similar way of $L_{sur}$ \textit{i.e.,}
\begin{align}
L_{sur}^{(ADM)}=\bar{L}_{sur}^{(ADM)}\Big|_{\kappa,\lambda\longrightarrow\kappa(t),\lambda(t)}-\frac{d}{dt}\Big[\frac{r^2\sin\theta\dot{\lambda}_i(\p g/\p\lambda_i)}{\sqrt{f}g^{3/2}}\Big]~. \label{LADMDIFF}
\end{align}

The expression of $L_{sur}^{(ADM)}$ is provided in Eq. \eqref{LSURADM}. Now, the metric \eqref{SSSFG} is static with no shift. Therefore, as we have mentioned earlier, all the components of $K_{ab}$ and its trace vanishes (\textit{i.e.} $K_{ab}=0$ and $K=0$). As a result, $\bar{L}_{sur}^{(ADM)}$ is given as $\bar{L}_{sur}^{(ADM)}=-2\p_{\alpha}(\g a^{\alpha})$ (also, $\bar{L}_{sur}^{(ADM)}|_{\kappa,\lambda\longrightarrow\kappa(t),\lambda(t)}=-2\p_{\alpha}(\g a^{\alpha})|_{\kappa,\lambda\longrightarrow\kappa(t),\lambda(t)}$). Now, for metric \eqref{SSTFG}, the only non-vanishing component of $K_{ab}$ is $K_{rr}$. Therefore, when the parameters are promoted to variables, $K$ does not vanish any more. Therefore, from \eqref{LSURADM} and \eqref{LADMDIFF}, we obtain
\begin{align}
\frac{d}{dt}\Big[\frac{r^2\sin\theta\dot{\lambda}_i(\p g/\p\lambda_i)}{\sqrt{f}g^{3/2}}\Big]=2\g\nabla_i(Kn^i)~, \label{RELATION}
\end{align}
which we have used in \eqref{DIFFEHLAG2}.
We also have shown how the Einstein-Hilbert Lagrangian is decomposed in terms of $G_{ab}n^an^b$ and $R_{ab}n^an^b$ in \eqref{DECOMP3}. We have mentioned that this decomposition, in static spacetime, enables us to interpret the Einstein-Hilbert  action as the free energy of the spacetime, where the integration over $G_{ab}n^an^b$ contributes as $\beta E$ and the integration over $R_{ab}n^an^b$ contributes as the entropy. Therefore, it motivates us to ask what happens in this decomposition of Einstein-Hilbert action if we promote the parameters to the time-dependent variables for a general SSS spacetime.

We find that $G_{ab}n^an^b$ does not change if we promote the constant parameters to variables in a SSS spacetime \textit{i.e.,}
\begin{align}
G_{ab}n^an^b=\bar{G}_{ab}\bar n^a \bar n^b\Big|_{\kappa,\lambda\longrightarrow\kappa(t),\lambda(t)}~,
\end{align}
and following the same argument as earlier we obtain 
\begin{align}
R_{ab}n^an^b=\bar{R}_{ab}\bar n^a\bar n^b\Big|_{\kappa,\lambda\longrightarrow\kappa(t),\lambda(t)}+\frac{d}{dt}\Big[\frac{r^2\sin\theta\dot{\lambda}_i(\p g/\p\lambda_i)}{2\sqrt{f}g^{3/2}}\Big]~.
\end{align}
 Thus, while interpreting Einstein-Hilbert action as the free energy of SSS spacetime, the term which is interpreted as $\beta E$, will remain invariant but, the term which is identified as the entropy, will change when the constant parameters are promoted to time-dependent variables.
 
 Let us now summarize all the results which we have obtained so far. The Einstein-Hilbert action can be decomposed in several forms. Each of the forms have different significances. Interestingly, for a general SSS spacetime, if we promote the constant parameters to time-dependent variables, one component in each of the decompositions remain invariant (such as $L_{quad}$, $L_{ADM}$ and $2G_{ab}n^an^b$). On the contrary, the other component in each decompositions (and the Einstein-Hilbert action itself) differ by the same total time-derivative term. The total-time derivative can be identified as $-2\g\nabla_i(Kn^i)$, which is equivalent to the negative of GHY surface term (considering the $16\pi$ factor properly) on a spacelike hyper-surface. 
 
 Once again, let us note that the difference in the expression of $\g R$, $L_{sur}$, $L_{sur}^{(ADM)}$ and $R_{ab}n^an^b$ appears only when the parameters of $g(r, \lambda)$ (of metric \eqref{SSSFG}) are promoted to time-dependent variables. If the parameters of $f(r, \kappa)$ are changed to time-dependent variables and the parameters of $g(r, \lambda)$ are left unchanged, there will be no difference in the expression of $\g R$, $L_{sur}$, $L_{sur}^{(ADM)}$ and $R_{ab}n^an^b$ as well. In fact, it can generally be proved that for \textit{any diagonal metric} (say described by the coordinates \{$t, x, y, z$\}) if the parameters of any one of the spatial metric components (\textit{i.e.} one among $g_{xx}$, $g_{yy}$ and $g_{zz}$) are promoted to time-dependent variables, it can be shown that $L_{bulk}$ and $L_{bulk}^{(ADM)}$ remain unchanged, while $\g R$, $L_{sur}$ and $L_{sur}^{(ADM)}$  change by a total time-derivative term. If the parameters in $g_{tt}$ are promoted to time-dependent variables and the parameters of the spatial components of the metric are left unchanged, the Einstein-Hilbert Lagrangian as well as the various components of decomposition ($L_{bulk}$, $L_{bulk}^{(ADM)}$, $L_{sur}$, $L_{sur}^{(ADM)}$ \textit{etc.}) remains unchanged. Therefore, if we consider the metric of the following form, 
\begin{align}
d\bar s^2=-f(x,\lambda)dt^2+dx^2+dy^2+dz^2~, \label{SSSF}
\end{align}
and promote $\lambda$'s to $\lambda(t)$'s it can be shown that neither $\g R$ changes nor any component of the decomposition (such as $L_{bulk}$, $L_{bulk}^{(ADM)}$, $\g R$, $L_{sur}$, $L_{sur}^{(ADM)}$ \textit{etc.}) remains unchanged (in this particular case, $G_{ab}n^an^b=0$ and $R=-2R_{ab}n^an^b$, which remain unchanged when $\lambda\longrightarrow\lambda(t)$). Note that when $f(x,\lambda)=(1+\alpha x)^2$ or $f(x,\lambda)=\alpha^2x^2$, it corresponds to the Rindler metric with vanishing Ricci-scalar. As the above analysis suggests, one can check that if one replaces the constant $\alpha$ by a time dependent function $\alpha(t)$, all the aforementioned comments will be valid and the spacetime still remains flat.

We particularly mention about the special case of SSS metric \eqref{SSSFG} when $f(r,\kappa)=g(r, \lambda)$. In that case, the metric is given as
\begin{align}
d\bar{s}^2= -f(r, \lambda)dt^2+\frac{dr^2}{f(r,\lambda)}+r^2d\Omega^2~. \label{SSSFF}
\end{align}
 Interestingly, for static metrics \eqref{SSSF} and \eqref{SSSFF}, the Einstein-Hilbert action is itself a total-derivative term. For metric \eqref{SSSF}, the expression of Einstein-Hilbert Lagrangian is given as $16\pi\sqrt{-\bar{g}}\bar{L}=\sqrt{-\bar{g}}\bar{R}=-d/dx(f'/\sqrt{f})$ and for metric \eqref{SSSFF} $\sqrt{-\bar{g}}\bar{R}=d^2/dr^2(r^2\sin\theta(1-f))$. Therefore, for these metrics, if we dynamically upgrade $\lambda$'s to time dependent as well as space-dependent function (\textit{i.e.,} $\lambda(t,x)$ for metric \eqref{SSSF} and $\lambda(t,r)$ for metric \eqref{SSSFF} respectively), the Einstein-Hilbert action will be modified by extra space-derivative for metric \eqref{SSSF} and by extra time-derivative plus a space-derivative for metric \eqref{SSSFF}.
 
 
\section{Obliterating the time-dependence from the metric parameters: Back to square one?} \label{OBLIT}
In several ways the geometry of spacetime can be changed significantly by a little modification in the metric. Dynamic upgradation of constant parameters in the metric is one such way, which we have discussed thoroughly in the previous section for the SSS spacetime. Another possibility can be the addition of extra parameters in the metric. In both the cases, the spacetime geometry will be different from the original spacetime. Then the question arises, if we go from the transformed spacetime to the original one (by obliterating the time-dependence from the metric parameters or by setting the extra parameters zero), does the geometry of the transformed spacetime reduces to the geometry of the original one? We investigate it by studying the properties of the Ricci-scalar.



We start this analysis with a simple example, which has been studied earlier in the literature \cite{Padmanabhan:2004fq,Padmanabhan:2003dc}.  We take the metric \eqref{SSSF} with $f(x,\lambda)=\epsilon^2+a^2x^2$. When $a=0$, it corresponds to the flat spacetime. However, when $\epsilon=0$, it corresponds to the flat spacetime in Rindler coordinates, which has horizon at $x=0$ and $R=0$ everywhere. On the contrary, if we account $\epsilon^2$ term in $f(x,\lambda)$ the corresponding expression of the Ricci-scalar is given as 
\begin{align}
R=-\frac{2\epsilon^2a^2}{(\epsilon^2+a^2x^2)^2}~. \label{RICCIEX1}
\end{align}
Now, at $\epsilon\longrightarrow 0$ limit, one expects to return back to the flat spacetime with $R=0$. However, the expression of Ricci-scalar in \eqref{RICCIEX1} suggests that for $\epsilon\longrightarrow 0$ one obtains 
\begin{align}
\lim_{\epsilon\rightarrow0}R=-2\delta(x^2)~,
\end{align}
which means the Ricci-scalar vanishes everywhere except on the horizon. This phenomenon has been mentioned as the concentration of curvature (for details about this phenomenon and its resemblance with an electromagnetic phenomenon, see \cite{Padmanabhan:2004fq,Padmanabhan:2003dc}).

\begin{figure}[h]
\centering
\includegraphics[width=\textwidth]{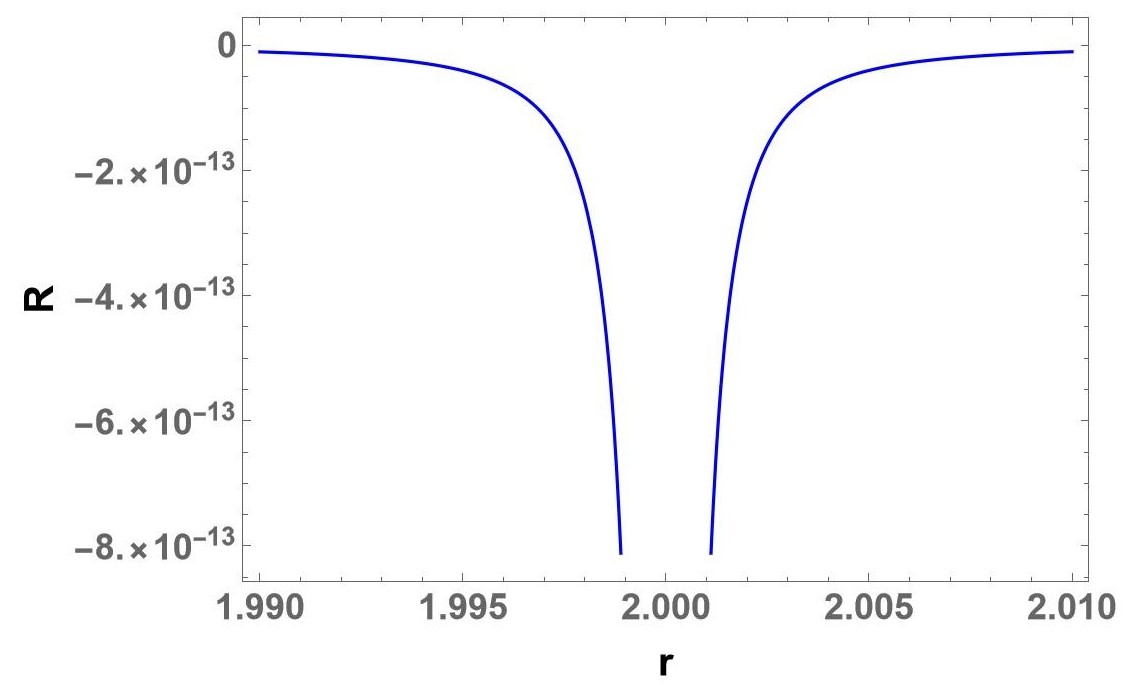}\caption{Behaviour of Ricci scalar $R$ (given by Eq. \eqref{RSWCHEPSILON}) near the horizon. For simplicity, we have assumed $M=1$ and $\epsilon=10^{-10}$~.}
\label{FIGRSWCHEPSILON}
\end{figure}

There are other such examples, which are not yet highlighted in literature. Which we have discussed in the following.
In Schwarzschild metric, one can introduce smallness parameter and write the spacetime metric as
\begin{align}
ds^2=-\Big(1-\frac{2M}{r}+\epsilon^2\Big)dt^2+\frac{dr^2}{\Big(1-\frac{2M}{r}-\epsilon^2\Big)}+r^2d\Omega^2~. \label{SWCHEPSILON}
\end{align}
Note that after introducing the parameter $\epsilon^2$, $g_{tt}\neq-g^{rr}$ in the metric \eqref{SWCHEPSILON}\footnote{If we consider $-g_{tt}=g^{rr}=1-2M/r+\epsilon^2$, we obtain $R=-2\epsilon^2/r^2$, which immediately vanishes when $\epsilon\rightarrow0$ and, therefore, do not have any diverging behaviour near the horizon.}. However, when $\epsilon=0$, the above metric \eqref{SWCHEPSILON} boils down to the usual Schwarzschild one, in which case $R=0$. However, the expression of scalar curvature corresponding to the metric \eqref{SWCHEPSILON} is given as
\begin{align}
R=\frac{2\epsilon^2\Big[\Big(r(1+\epsilon^2)-2M\Big)^2-2M^2\Big]}{r^2\Big(r(1+\epsilon^2)-2M\Big)^2}~. \label{RSWCHEPSILON}
\end{align}
It can be proved that
\begin{align}
\lim_{\epsilon\rightarrow0}R=\frac{2\epsilon^2}{r^2}-\Big(\frac{2M}{r}\Big)^2\delta(r-2M)~. 
\end{align}
Therefore, only last term becomes significant near $r=2M$. When $\epsilon\longrightarrow0$, one finds that $R\longrightarrow0$ everywhere except very near to the horizon. In figure \ref{FIGRSWCHEPSILON}, we describe the behaviour of the Ricci scalar R(as given in \eqref{RSWCHEPSILON}) near the horizon $r_H=2M$.

 In \cite{Padmanabhan:2003dc}, it has been mentioned that any metric, which can be approximated to the Rindler metric near the horizon, can have a diverging Ricci-scalar near the horizon when the smallness parameter goes to zero. This statement can be verified for Schwarzschild metric. It can be shown that, if we define $l=r-r_H$ (where $r_H=2M$ is the horizon), in the limit $l<<r_H$ (\textit{i.e.} near the horizon) $f(r)\rightarrow h(l)$ 
 where $h(l)=2al$, with $a=f'(r_H)/2=1/2r_H$. Thus, in this limit, the Schwarzschild metric takes the form of Rindler metric in the $t-l$ coordinates. Therefore, adopting the similar prescription to that of \cite{Padmanabhan:2003dc} (where, $h(l)$ is deformed as $h(l)\longrightarrow\sqrt{\epsilon^2+h(l)^2}$, see Eq. (19) of \cite{Padmanabhan:2003dc}), here we deform $f(r)$ as $f(r)\longrightarrow\sqrt{\epsilon^2+f(r)^2}$. Thus, the asymptotic Schwarzschild metric looks
 \begin{align}
 d\bar s^2=-\sqrt{\epsilon^2+(1-\frac{2M}{r})^2}dt^2+\frac{dr^2}{\sqrt{\epsilon^2+(1-\frac{2M}{r})^2}}+r^2d\Omega^2~,
 \end{align}
 which reduces to the Schwarzschild metric in the limit $\epsilon\rightarrow0$. The Ricci scalar of the above metric is given as
 \begin{align}
 R=\frac{-4 M^2 \epsilon ^2}{r^4\Big((1-\frac{2M}{r})^2+\epsilon^2\Big)^{\frac{3}{2}}}-\frac{2\Big(1-\frac{2M}{r}+\epsilon^2\Big)}{r^2 \sqrt{\left(1-\frac{2 M}{r}\right)^2+\epsilon ^2}}+\frac{2}{r^2}~. \label{RICCISCHRIND}
 \end{align}
 It can be checked that, away from the horizon (\textit{i.e.} for $r\neq 2M$), the Ricci scalar vanishes in the limit $\epsilon\rightarrow 0$. At the horizon (\textit{i.e.} $r= 2M$), it can be checked that the behaviour of $R$ is given as
 \begin{align}
 \lim_{\epsilon\rightarrow0}R=-\frac{1}{4M^2}\delta(1-\frac{2M}{r})-\frac{\epsilon}{2M^2}+\frac{1}{2M^2}~.
 \end{align}
 Thus, in the limit $r=2M$, the first term of Eq. \eqref{RICCISCHRIND} is the leading order term, which diverges near the horizon. This justifies the statement of \cite{Padmanabhan:2003dc}. In addition, this also suggests that there can be many possible ways to obtain divergence in the Ricci scalar for the same metric, deformed by smallness parameters in different ways. 
 


\begin{figure}[h]
\centering
\includegraphics[width=\textwidth]{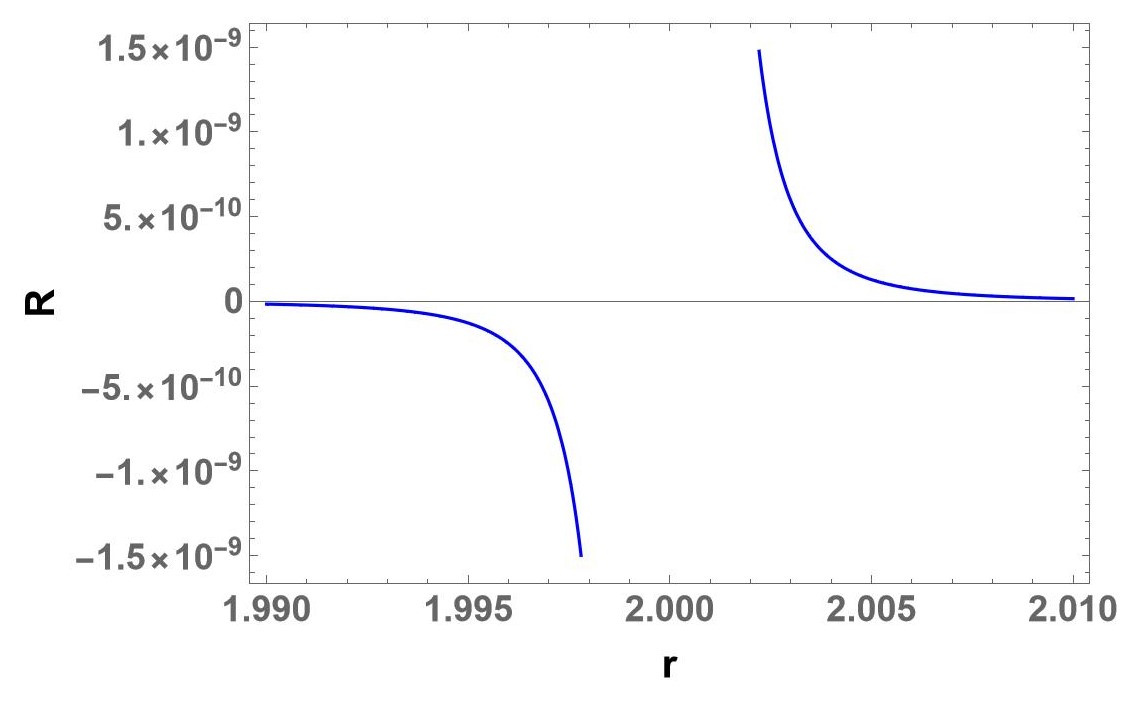}
\caption{Behaviour of Ricci scalar R (given by Eq. \eqref{FORPLOTR}) near the horizon. For simplicity, we have assumed $M=1$, $t=1$ and $\epsilon=10^{-10}$~.}
\label{FIGRSWCHTDEP}
\end{figure}

\vskip 5mm
Now we come back to the time-dependent cases. We found that the divergence in the Ricci scalar also happens when we remove the time-dependence of the parameters. For instance, the Schwatzschild spacetime is Ricci flat. If we promote the parameter $M\longrightarrow M(t)$, the expression of Ricci-scalar is given a total derivative term as given by Eq. \eqref{TRSCH}, which can be expressed in a more convenient form as
\begin{align}
R=\frac{8r\dot{M}(t)^2}{(r-2M(t))^3}+\frac{2r\ddot{M}(t)}{(r-2M(t))^2}~. \label{RTDEPSCH}
\end{align}
For simplicity, first we consider 
\begin{align}
M(t)=M\exp(\epsilon t)~, \label{MT}
\end{align}
where $M$ is independent of $t$ and $\epsilon$ is the smallness parameter. If $\epsilon=0$, the metric will boil down to the Schwarzschild one. If one considers $M(t)$ of the form given in \eqref{MT}, the expression of the Ricci-scalar will be given as
\begin{align}
R=\frac{8r\epsilon^2M^2\exp(\epsilon t)}{(r-2M\exp(\epsilon t))^3}+\frac{2r\epsilon^2M\exp(\epsilon t)}{(r-2M\exp(\epsilon t))^2}~. \label{FORPLOTR}
\end{align}
 When $r\neq 2M$, one obtains $R=0$ for $\epsilon=0$~. However, near the horizon of static Schwarzschild black hole (\textit{i.e.} $r_{H}=2M$), one obtains
 \begin{align}
 \lim_{\epsilon\rightarrow0} R\sim\frac{\mathcal{O}(\epsilon^2)}{\Big(r-2M+\mathcal{O}(\epsilon)\Big)^3} \label{LIMITR}
 \end{align}
 Therefore, near the horizon, the Ricci-scalar appears to be a diverging quantity. However, this divergence on the horizon is not exactly similar as it has been the case in Eq. \eqref{RICCIEX1}. This is because, the denominator in Eq. \eqref{LIMITR} changes its sign on the point $r=2M$ for $\epsilon\longrightarrow0$. In this case, the Ricci-scalar shows infinite discontinuity on the horizon. In figure \ref{FIGRSWCHTDEP} we have shown the behaviour of the Ricci scalar (as given by Eq. \eqref{FORPLOTR}) near the horizon.

Although the infinite discontinuity of curvature has been shown for $M(t)$ having a particular (exponential) time dependence (as shown in Eq. \eqref{MT}), this phenomenon is not limited to that time-dependence only. In general, the smallness parameter $\epsilon$ should be associated with time in such a way that when $\epsilon=0$, $M(\epsilon=0,t)=M$, where $M$ is the mass of the static Schwarzschild black hole. Also, we must have $M^{(n)} \leq\mathcal{O}(\epsilon)$, $\forall n\in\mathbb{N}$; where $M^{(n)}$ corresponds to the $n$-th order derivative of $M(t)$ with respect to $t$. This is because, when $\epsilon=0$, $M^{(n)}$ should vanish.

 
  Now, for a general $M(t)\equiv M(\epsilon,t)$ can be expanded in terms of Maclaurin's series as
\begin{align}
M(\epsilon, t)=M+\sum_{i=1}^{\infty}\epsilon^i m_{i}(t)~, \label{MMaclurin}
\end{align}
where 
\begin{align}
m_{k}(t)=\frac{1}{k!}\frac{\p^k M(\epsilon, t)}{\p\epsilon^k}\Big|_{\epsilon=0}
\end{align}
Therefore, from \eqref{MMaclurin}, we obtain 
\begin{align}
\dot M(\epsilon, t)=\epsilon\dot m_1(t)+\mathcal{O}(\epsilon^2)+...~,\label{MDOT}
\end{align}
and 
\begin{align}
\ddot M(\epsilon, t)=\epsilon\ddot m_1(t)+\mathcal{O}(\epsilon^2)+....\label{MDOTDOT}
\end{align}
Note that \eqref{MDOT} and \eqref{MDOTDOT} suggest that $M^{(n)}$ vanishes at the limit $\epsilon\rightarrow 0$.
Therefore, generally, we have $\dot M(\epsilon, t)\sim \mathcal{O} (\epsilon)$  and the expression of $R$ (given in \eqref{FORPLOTR}) boils down to the expression of \eqref{LIMITR}.
Hence, for a general $M(\epsilon, t)$, it can be concluded that the near-horizon expression of the Ricci-scalar is given by the form \eqref{LIMITR} and the near horizon behaviour of the Ricci-scalar is given by the figure \ref{FIGRSWCHTDEP}. 


 This phenomena can be generalized to the spherically symmetric metrics. Consider the metric \eqref{SSSFF}. When $\lambda\longrightarrow\lambda(t)$, the scalar curvature is obtained as 
\begin{align}
R=\bar{R}\Big|_{\lambda\longrightarrow\lambda(t)}+R'~,
\end{align}
where $\bar{R}$ corresponds to the Ricci-scalar of static and spherically symmetric metric \eqref{SSSFF} and 
\begin{align}
R'=-\frac{d}{dt}\Big[\frac{1}{f^2}\Big(\dot{\lambda}\frac{\p f}{\p\lambda}\Big)\Big]=-\frac{\ddot{\lambda}\frac{\p f}{\p\lambda}+\dot{\lambda}^2\frac{\p^2f}{\p\lambda^2}}{f(r, \lambda(t))^2}+\frac{2\dot{\lambda}^2(\frac{\p f}{\p\lambda})^2}{f(r, \lambda(t))^3}~.\label{rprime}
\end{align}
Now, we consider $\lambda(t)$ contains the smallness parameter such a way that for $\epsilon=0$, $\lambda(t)\equiv\lambda(t, \epsilon)=\lambda$ (where $\lambda$ is the constant parameter before the dynamic upgradation) and $\lambda^{(n)}(t)\leq\mathcal{O}(\epsilon)$, $\forall n\in\mathbb{N}$; where $\lambda^{(n)}(t)$ corresponds to the $n$-th order derivative of $\lambda(t)$ with respect to $t$. Now, one can make Maclaurin series expansion of $f(r, \lambda(\epsilon, t))$ as
\begin{align}
f(r, \lambda(\epsilon, t))= f(r, \lambda)+\epsilon\Big(\frac{\p f}{\p\lambda}\frac{\p\lambda(\epsilon,t)}{\p\epsilon}\Big)_{\epsilon=0}+\mathcal{O}(\epsilon^2)+..
\end{align}
Again, as we did in \eqref{MMaclurin}, we can make Maclaurin series expansion of $\lambda(t)$ and generally show that $\dot{\lambda}\sim \mathcal{O}(\epsilon)$. As a result, we obtain that for $\epsilon\longrightarrow 0$, $R'$ vanishes everywhere except near the horizon (defined by $f(r_{H},\lambda)=0$), where the expression is given as (the last term of \eqref{rprime} becomes the leading order term)
\begin{align}
\lim_{\epsilon\rightarrow0}R'\sim \frac{\mathcal{O}(\epsilon^2)}{(f(r, \lambda)+\mathcal{O}(\epsilon))^3}~,
\end{align}
which is a diverging quantity near the horizon. Also, since $f(r, \lambda)$ changes its sign at $r=r_H$, the near horizon behaviour of $R'$ should be the same as described in figure \ref{FIGRSWCHTDEP}.


For a general SSS metric of the form given by \eqref{SSSFG}, the situation is a little different. When $\lambda$ are $\kappa$ are promoted to $\lambda(t)$ are $\kappa(t)$, the expression of $\g R$ is given by $\g R=\g \bar{R}|_{\kappa,\lambda\longrightarrow\kappa(t),\lambda(t)}+\g R'$, where $\bar{R}$ is given by \eqref{RICCISSSFG} and $\g R'$ is defined in \eqref{DIFFEHLAG}, which can be further expressed in a more convenient form as
\begin{align}
\g R'=-\frac{d}{dt}\Big[\frac{r^2\sin\theta\dot{\lambda}_i(\p g/\p\lambda_i)}{\sqrt{f}g^{3/2}}\Big]=r^2\sin\theta\Big[\frac{\dot{\lambda}\dot{\kappa}}{2(fg)^{\frac{3}{2}}}\Big(\frac{\p f}{\p\kappa}\Big)\Big(\frac{\p g}{\p\lambda}\Big)+\frac{3}{2}\frac{\dot{\lambda}^2}{\sqrt{f}g^{\frac{5}{2}}}\Big(\frac{\p g}{\p\lambda}\Big)^2
\nonumber
\\
-\frac{1}{\sqrt{f}g^{\frac{3}{2}}}\Big(\ddot{\lambda}\frac{\p g}{\p\lambda}+\dot{\lambda}^2\frac{\p^2g}{\p\lambda^2}\Big)\Big]~.
\end{align}
For simplicity, we consider $\lambda(t)=\lambda\exp(\epsilon t)$ and $\kappa(t)=\kappa\exp(\epsilon t)$. With this, we can show, for $\epsilon\longrightarrow 0$, $R'\longrightarrow0$  everywhere except those places where $g(r, \lambda)=0$ and/or $f(r, \kappa)=0$.

 For the static \& spherically symmetric metric \eqref{SSSFG}, $g(r, \lambda)=0$ corresponds to the \textbf{apparent horizon} (defined by $g^{rr}|_{r=r_{AH}}=g(r_{AH}, \lambda)=0$, where $r_{AH}$ stands for the radius of the apparent horizon). Near $r=r_{AH}$, we can again obtain the Maclaurin series of $g(r, \lambda(t))$ and show that $\g R'$ diverges as
\begin{align}
\g R'\sim \frac{\mathcal{O}(\epsilon^2)}{(g(r,\lambda)+\mathcal{O}(\epsilon))^{\frac{5}{2}}}~,
\end{align} 
 where we have considered the fact that $f(r_{AH}, \kappa)$ is non-zero finite. Note that in this case, $R'$ alone becomes a non-zero finite quantity \textit{i.e.} $\lim_{r\rightarrow r_H}R'\sim\mathcal{O}(1)$ (again, considering $f(r_{AH}, \kappa)$ is non-zero finite). Therefore, we observe that if we put the smallness parameter $\epsilon\rightarrow$, the Ricci-scalar do not agree with its static value $\bar{R}$ on the apparent horizon and there we found the divergence of Einstein-Hilbert Lagrangian.
 
 For metric \eqref{SSSFG}, $f(r, \kappa)=0$ corresponds to the \textbf{Killing horizon} (defined by $g_{tt}|_{r=r_{KH}}=f(r_{KH}, \kappa)=0$, where $r_{KH}$ stands for the radius of the Killing horizon). In that case, for $\lambda(t)=\lambda\exp(\epsilon t)$ and $\kappa(t)=\kappa\exp(\epsilon t)$ one obtains $R'\sim\mathcal{O}(1)$ for $r=r_{KH}$ (considering $g(r_{KH}, \lambda)\neq 0$). As a result $\g R'$ vanishes. Therefore, again we found that, if we put the smallness parameter $\epsilon=0$, the metric reduces to the static one. But the Ricci-scalar do not agree with its static value $\bar{R}$. However, the Einstein-Hilbert Lagrangian, unlike on the apparent horizon, agree to its static value on the Killing horizon.
 
 Thus, in this section, we have found that even after obliterating the time dependence (by setting the smallness parameter associated with time to zero), the curvature and the EH action does not (usually) boil down to its static value. Instead, it becomes singular near the horizon. Before that, we have shown that even in static cases (Rindler and Schwarzschild metric), if one deform the metric with a smallness parameter, the curvature and the EH action becomes a Dirac-delta function on the horizon. This implies a permanent topological change which is irreversible. The cause of such peculiar behaviour is not known yet and should in investigated in future.
 
\section{Coordinate transformation vs dynamic upgradation /reduction of parameters} \label{COPMA}
The coordinate transformation does not change the geometry of the spacetime. On the contrary, as we have discussed earlier, the spacetime geometry indeed gets  changed if one promotes the constant parameters to variables. Therefore, for a same metric expressed in different coordinates, if we promote the constant parameters to time-dependent variables, the change of Einstein-Hilbert action and its various components are not the same. In the following, we provide some simple examples to demonstrate this in a more detail.

 A flat metric in Rindler coordinate is given as
\begin{align}
d\bar s^2=-(1+ax)^2dt^2+dx^2+dy^2+dz^2~. \label{RIND1}
\end{align}
 By a coordinate transformation $(1+ax)=a\bar x$, the metric \eqref{RIND1} can be written alternatively as
 \begin{align}
 d\bar s^2=-a^2\bar x^2dt^2+d\bar x^2+dy^2+dz^2~. \label{RIND2}
 \end{align}
One can make another coordinate transformation $(1+ax)=\sqrt{2al}$ in metric \eqref{RIND1} and obtain
\begin{align}
d\bar s^2=-2aldt^2+\frac{dl^2}{2al}+dy^2+dz^2~. \label{RIND3}
\end{align}
Among these three expressions of the Rindler metric, the first two, provided in Eqs. \eqref{RIND1} and \eqref{RIND2}, corresponds to the metric \eqref{SSSF}. Therefore, from our earlier analysis, we obtain that the spacetime is still flat if we promote $a\longrightarrow a(t)$ in \eqref{RIND1} and \eqref{RIND2}. In addition, various components of the decompositions of Einstein-Hilbert action will remain unaffected for $a\longrightarrow a(t)$. Since, the metrics \eqref{RIND1} and \eqref{RIND2} are flat, even after promoting $a\longrightarrow a(t)$, we provide the transformation relation of those coordinates with the Minkowski one (say, described by the coordinates $T$, $X$, $Y$, $Z$).

The metric \eqref{RIND1}, with $a\longrightarrow a(t)$, is related to the Minkowski coordinates as follows (see project 3.4 of \cite{gravitation})
\begin{align}
X=\int'\sinh\chi(t)dt+x\cosh\chi(t)~,
\nonumber
\\
T=\int'\cosh\chi(t)dt+x\sinh\chi(t)~. \label{TRRIN1}
\end{align}
Also, $a(t)$ is given by $a(t)=d\chi/dt$~. When $a$ is constant (\textit{i.e.} in metric \eqref{RIND1}), the transformation relations can simply be obtained from \eqref{TRRIN1} as $X=(1+x)\cosh(at)$ and $T=(1+x)\sinh(at)$~.

The metric \eqref{RIND2}, with $a\longrightarrow a(t)$, is related to the Minkowski coordinates as 
\begin{align}
X=\bar x\cosh\chi(t)~,
\nonumber
\\
T=\bar x\sinh\chi(t)~, \label{TRRIN2}
\end{align}
and $a(t)=d\chi/dt$~. When $a$ is constant, the transformation relation can simply be obtained from \eqref{TRRIN2} as $X=\bar x\cosh(at)$ and $T=\bar x\sinh(at)$~.

 On the other hand, the metric \eqref{RIND3} is Ricci-flat. Now, if we promote $a\longrightarrow a(t)$, the spacetime will no longer remain flat. One can obtain the expression of the Ricci scalar as
\begin{align}
R=\frac{2 \dot a(t)^2-a(t) \ddot a(t)}{2 l a(t)^3}=-\frac{\p}{\p t}\Big(\frac{\dot a(t)}{2la(t)^2}\Big)~. \label{RICCIRINDLERT}
\end{align}
In addition, it can be shown that there will be  a divergence of the Ricci-scalar (or EH action) near the horizon if the time dependence is removed by incorporating the smallness parameter. For example, if we consider $a(t)$ is given as $a(t)=a\exp[\epsilon t]$, the expression of the Ricci scalar (given by the Eq. \eqref{RICCIRINDLERT}), at the limit $\epsilon\longrightarrow0$, can be obtained as
\begin{align}
\lim_{\epsilon\rightarrow0}R=\frac{\epsilon^2}{2la},
\end{align} 
which vanishes everywhere in the limit $\epsilon\longrightarrow0$ except very near the horizon $0\leq l<\epsilon^2$ (note that for the metric \eqref{RIND3}, the horizon is at $l=0$).

Thus, in the above, we have argued how the geometry of the same metric, expressed in different coordinates, changes after promoting the constant parameters to time-dependent variables. Note that in the above three metrics \eqref{RIND1}, \eqref{RIND2} and \eqref{RIND3} the time coordinate does not transform. Nevertheless, if the constant parameter $a$ is promoted to $a(t)$, the geometry of the spacetime changes in different way in different coordinates.

\begin{figure}[h]
\centering
\includegraphics[width=\textwidth]{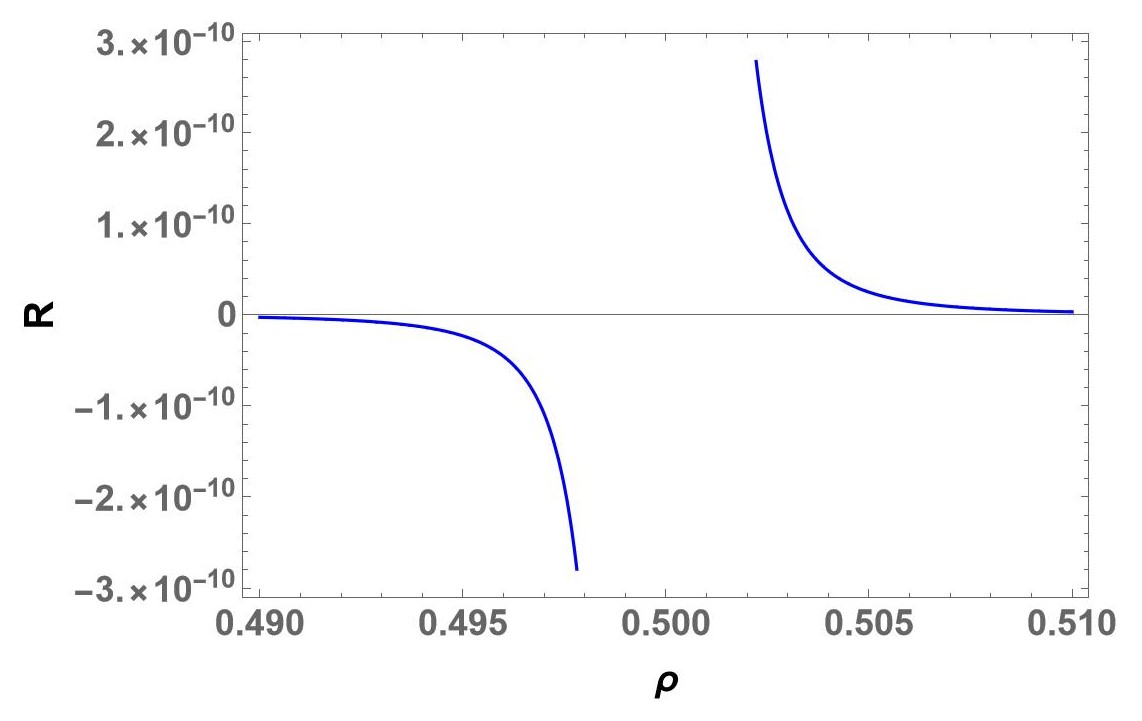}
\caption{Behaviour of Ricci scalar R (given by Eq. \eqref{RSCHISO} with $M(t)$ given by Eq. \eqref{MT}) near the horizon. For simplicity, we have assumed $M=1$, $t=1$ and $\epsilon=10^{-10}$~.}
\label{FIGISO}
\end{figure}

We provide another example in the following. We have mentioned earlier how the Ricci-scalar changes in Schwarzschild metric, when we promote $M\longrightarrow M(t)$. Also we have mentioned how the curvature gets concentrated on the horizon if we remove the time-dependence using a smallness parameter. Now, the Schwarzschild metric can be expressed in several other coordinates. For example, we can express the Schwarzschild metric in the \textit{isotropic coordinate} as
\begin{align}
d\bar s^2=-\Bigg(\frac{1-\frac{M}{2\rho}}{1+\frac{M}{2\rho}}\Bigg)^2dt^2+\bigg(1+\frac{M}{2\rho}\bigg)^4(d\rho^2+\rho^2d\Omega^2)~, \label{SCHISO}
\end{align}
where the radial coordinate of Schwarzschild coordinate is connected to the same in isotropic coordinate as $r=\rho(1+(M/2\rho))^2$~. The Ricci-scalar of the metric \eqref{SCHISO}, of course, vanishes. However, if we now promote $M\longrightarrow M(t)$ in the metric \eqref{SCHISO}, the expression of the Ricci-scalar turns out as
\begin{align}
R=\frac{12[\dot{M}^2(10\rho-3M(t))+\ddot{M}(4\rho^2-M(t)^2)]}{(2\rho-M(t))^3}~. \label{RSCHISO}
\end{align}
Note that the expression of the Ricci-scalar, as provided in \eqref{RSCHISO}, does not boil down to the expression provided in \eqref{RTDEPSCH} using the transformation relation between $\rho$ and $r$.  The horizon of the static metric \eqref{SCHISO} is located as $\rho_H=M/2$.
If one takes the form of $M(t)$ of the form \eqref{MT}, one can again show that he Ricci-scalar everywhere when $\epsilon\rightarrow 0$ except on the horizon, where it diverges as $R\sim\mathcal{O}(1/\epsilon)$~. The near-horizon behaviour of the Ricci-scalar is shown in figure \ref{FIGISO}.
As figure \ref{FIGISO} shows, the near-horizon behaviour of the Ricci-scalar corresponding to the metric \eqref{SCHISO} with $M\longrightarrow M(t)$ is similar to that of the metric \eqref{TDEPSCH}.

The Schwarzschild metric can also be expressed in \textit{Painleve-Gullstrand coordinates} as
\begin{align}
d\bar s^2=-\Big(1-\frac{2M}{r}\Big)dT^2-2\sqrt{\frac{2M}{R}}dTdr+dr^2+r^2d\Omega^2~, \label{SCHPAIN}
\end{align}
where $T$ is connected to the Schwarzschild time coordinate $t$ as
\begin{align}
T=t+4M\Bigg(\sqrt{\frac{r}{2M}}+\frac{1}{2}\ln\Bigg|\frac{\sqrt{r/2M}-1}{\sqrt{r/2M}+1}\Bigg|\Bigg)~.
\end{align}
Again, the Ricci-scalar corresponding to the metric \eqref{SCHPAIN} vanishes. However, if we promote $M\longrightarrow M(T)$, the corresponding expression of the Ricci-scalar is given as 
\begin{align}
R=\frac{3\dot{M}}{r^2\sqrt{\frac{2M(T)}{r}}}~.\label{RSCHPAIN}
\end{align}
Here, $\dot{M}\equiv\p M/\p T$~. Now, if we consider $M(T)=M\exp(\epsilon T)$, we find that $R$ vanishes everywhere (even on the horizon $r=2M$) when $\epsilon\longrightarrow0$.

We end up this section by providing a more general analysis. A general SSS metric (of the form\eqref{SSSFF}) can be written in painleve-Gullstrand coordinates as
\begin{align}
d\bar s^2=-f(r,\lambda)dT^2-2\sqrt{1-f(r,\lambda)}dTdr+dr^2+r^2d\Omega^2~,\label{SSSFFPG}
\end{align}
where, $dt=dT+(\sqrt{1-f(r,\lambda)}/f(r,\lambda))dr$. Now, it is obvious that the expression of the Ricci scalars corresponding to the metrics \eqref{SSSFF} and \eqref{SSSFFPG} are the same. However, the interesting fact is that the quadratic part of the Einstein-Hilbert Lagrangian is also the same of the metrics \eqref{SSSFF} and \eqref{SSSFFPG}. Therefore, the surface part of the Lagrangian is also the same for the metrics \eqref{SSSFF} and \eqref{SSSFFPG}. We have earlier mentioned that the decomposition of the Einstein-Hilbert Lagrangian in terms of quadratic and the surface parts are not covariant one and the quadratic and the surface part are not covariant scalars. But, for any arbitrary static spacetime (described by the coordinates say \{$t$, $x^{\alpha}$\}) if we change the time coordinate by $t'=t+\bar f(x^{\alpha})$ then, in new coordinate system \{$t'$, $x^{\alpha}$\}, one can prove that the quadratic and the surface part of the Einstein-Hilbert Lagrangian remains invariant due to that specific coordinate transformation. The transformation from SSS spacetime \eqref{SSSFF} to the Painleve metric \eqref{SSSFFPG} is a special case of this result.
 
The expression of the Ricci scalar corresponding to the metric \eqref{SSSFFPG} is the same as of the metric \eqref{SSSFF}, which is given as 
\begin{align}
\sqrt{-\bar g^{(PG)}}\bar R^{(PG)}=\frac{d^2}{dr^2}\Big(r^2\sin\theta(1-f)\Big) \label{STSTICPG}
\end{align}
In the above equation, $PG$ stands for painleve-Gullstrand coordinates.
 Now, if we promote the constant parameters $\lambda$'s to $\lambda(T)$'s in \eqref{SSSFFPG}, the corresponding Ricci-scalar can be obtained as 
\begin{align}
\sqrt{-g^{(PG)}}R^{(PG)}=\sqrt{-\bar g^{(PG)}}\bar R^{(PG)}\Big|_{\lambda\longrightarrow\lambda(T)}+2\p_r\p_t\Big(\sqrt{-g^{(PG)}}\sqrt{1-f}\Big)~, \label{RPGTRANSFORM}
\end{align}
where $\p_r\equiv \p/\p r$ and $\p_t\equiv \p/\p t$.
Again, in Painleve-Gullstrand coordinates, the change in Einstein-Hilbert Lagrangian comes as a total derivative term (can be expressed as either total space or total space ($r$) derivative) if we promote the constant parameters to the time-dependent variables. Therefore, the expression on the RHS of Eq. \eqref{RSCHPAIN} can be expressed in terms of either total time or total space ($r$) derivative.

Note that although the change in the expression of the EH action appears to be a total derivative term after the dynamic upgradation of the constant parameters (like the SSS metric case), once the time-dependence is obliterated, the expression of EH action (or the Ricci-scalar) boils down to its earlier expression (given by Eq. \eqref{STSTICPG}) and there is no divergence in this case near the horizon. This can be shown from the following mathematical analysis. The extra term of \eqref{RPGTRANSFORM} can be written explicitly as
\begin{align}
2\p_r\p_t\Big(\sqrt{-g^{(PG)}}\sqrt{1-f}\Big)=-\frac{\dot\lambda_i\sin{\theta}}{\sqrt{1-f}}\Bigg[2r\frac{\p f}{\p\lambda_i}+r^2\frac{\p^2f}{\p r\p\lambda_i}+\frac{r^2}{2(1-f)}\frac{\p f}{\p\lambda_i}\frac{\p f}{\p r}\Bigg]~. \label{EXTRATRM}
\end{align}
In the above expression, the extra term is proportional to $\dot\lambda(t)$, which vanishes when we remove the time dependence (for example, if we consider $\lambda(t)=\lambda\exp{\epsilon t}$, $\dot\lambda(t)\sim \mathcal{O}(\epsilon)$). On the other hand, denominator is non-zero finite near the horizon. Therefore, at the limit $\epsilon\rightarrow 0$, we obtain $\sqrt{-g^{(PG)}}R^{(PG)}\longrightarrow\sqrt{-\bar g^{(PG)}}\bar R^{(PG)}$ and the extra term (of Eq. \eqref{EXTRATRM}) vanishes. This is true for Eq. \eqref{RSCHPAIN} as well \textit{i.e.} $R$ of Eq \eqref{RSCHPAIN} vanishes for $\epsilon\rightarrow 0$.

 There is another difference of SSS and the metric in Painleve-Gullstrand coordinates is the following. For SSS metric \eqref{SSSFF}, if we promote $\lambda\longrightarrow\lambda(t)$, earlier we have found that the quadratic part of the Einstein-Hilbert Lagrangian remains invariant. However, for the metric in Painleve-Gullstrand coordinates \eqref{SSSFFPG}, we obtain that the quadratic part of the Lagrangian is no more invariant when $\lambda\longrightarrow\lambda(T)$. It changes as follows
\begin{align}
L_{quad}^{(PG)}=\bar L_{quad}^{(PG)}\Big|_{\lambda\longrightarrow\lambda(T)}-2\frac{\p}{\p t}\Big[\frac{\sqrt{1-f}}{r}\Big]~.
\end{align}
Finally, we comment on the Kruskal coordinates. The Schwarzschild metric in Kruskal coordinates ($T$, $X$, $\theta$, $\phi$) are given as \cite{gravitation}
\begin{align}
d\bar{s}^2=\frac{32M^3}{r}\exp{(-r/2M)}(-dT^2+dX^2)+r^2 d\Omega^2~.
\end{align}
where $r$ is an implicit function of $X$ and $T$ determined by the following relation 
\begin{align}
\Big(\frac{r}{2M}-1\Big)\exp{(r/2M)}=X^2-T^2~.
\end{align}
Now, if we promote $M\rightarrow M(t)$, we obtain the expression of the Ricci-scalar as 
\begin{align}
R=\frac{1}{64 M(T)^6 r^2}\Bigg[M(T)^2 r^3 e^{\frac{r}{2 M(T)}} \left(-\p^2_Tr+\p^2_Xr+6 \ddot{M}\right)-2 \dot M^2 r^4 e^{\frac{r}{2 M(T)}}
\nonumber
\\
+M(T) r^3 e^{\frac{r}{2 M(T)}} \left(\ddot{M} r+2 \dot M \p_Tr-6 \dot M^2\right)+6 M(T)^3 r (\p_Tr^2-\p_Xr^2
\nonumber
\\
+r \left(\p^2_Tr-\p^2_Xr\right)) e^{\frac{r}{2 M(T)}}+128 M(T)^6\Bigg]
\end{align}
Since the denominator does not vanish when the time-dependence is removed by introducing smallness parameter, it can be said that the divergence of the Ricci-scalar does not happen in this case.

Thus, in this section, we see that for the same metric when it is expressed in different coordinates, they imply same geometry. But, after the promotion of constant parameters to time dependent variables, they imply different geometry. This is because, the dynamically upgraded metrics are no longer connected by a coordinate transformation.

 \section{Conclusions} \label{CONCLU}
 As it has been mentioned, the Einstein-Hilbert action is peculiar and different in many ways as compared to the actions of other fundamental theories. In the present paper, we have discussed the EH action and its different decompositions in great details and also have discussed some of the peculiar features of the EH action, which has not been reported earlier. Thus, the present paper can be viewed as a review of EH action along with some new findings.
 
 In section \ref{mathbg}, we have provided a concise review on various decomposition of the Einstein-Hilbert action and their underlying motivations. Here, we have discussed three ways of decomposing the EH action into the bulk and the surface parts and their connection with the gravitational thermodynamics. In addition, we also have discussed how the expressions of the bulk (or the surface) parts of these decompositions differ from one another for a given metric.
 
  In section \ref{DUP}, we have discussed how the expression of the Ricci-scalar (or the EH action) changes when the constant parameters of a static spacetime metric are promoted to time-dependent variables, which we call as dynamical upgradation of the constant parameters. We have found that for static and spherically symmetric metric, after the dynamical upgradation, the bulk parts of each decomposition remain invariant, whereas the surface part (of each decomposition) change by a total time-derivative. Therefore, the expression of EH action (or the Ricci-scalar), as a whole, changes by the same total time-derivative. As it has been noticed, the invariance of the bulk part in each decomposition, is a feature of spherically symmetric metric (\textit{i.e.} not general) as it no longer remains invariant for other coordinates (such as Painleve coordinate or in the isotropic coordinate). In this context, we also have obtained that for the metric of the form \eqref{SSSF}, if we dynamically upgrade the constant parameters (making them time-dependent variable), the expression of scalar curvature (or the EH action) does not change at all. This has been explicitly shown for the Rindler metrics \eqref{RIND1} and \eqref{RIND2}, where if we promote $a\longrightarrow a(t)$, the Ricci scalar still vanishes. On the other hand, for the Rindler metric \eqref{RIND3}, the dynamic upgradation $a\longrightarrow a(t)$ leads to the singular behaviour of the Ricci scalar when we remove the time-dependence introducing the smallness parameter. Furthermore, we have noticed that for the static and spherically symmetric spacetime metric, the dynamic upgradation of the parameters which are present in $g_{rr}$ (we have distinguished them as $\lambda_i$'s) are responsible for the change of the Ricci-scalar in terms of a total time-derivative. Whereas, the dynamic upgradation of the parameters present in $g_{tt}$, which we identify as $\kappa_i$'s does not alter the expression of the Ricci scalar or its decompositions.

 In section \ref{OBLIT} and \ref{COPMA}, we explore on the consequences after the removal of the time-dependence on the dynamically upgraded parameters. Also, we discuss how the dynamic upgradation results differently for the same metric expressed in different coordinates.  After the dynamic upgradation, we introduce a smallness parameter $\epsilon$ in such a way that when $\epsilon\longrightarrow 0$, the upgraded parameter reduces to its constant value. However, we find that in the limit $\epsilon\longrightarrow 0$, the curvature (or the EH Lagrangian) becomes singular near the horizon for a general spherically symmetric metric, Rindler metric (of form \eqref{RIND3}) and for Schwarzschild metric in isotropic coordinates. This means that  the change in geometry in the spacetime (due to the dynamic upgradation of the constant metric parameters) is permanent and not reversible even after setting the constant parameters to its constant value. We also have discussed a few examples in the static spacetime metrics, which are Ricci-flat (such as the Rindler metric, Schwarzschild metric). We deform those metrics by adding smallness parameters. Even in those cases, it was found that the expression of the Ricci-scalar does not vanish after setting the smallness parameter to zero and we obtain the singular behaviour of the scalar curvature. In addition, we have shown that for the Schwarzschild metric (which is also true for a general SSS metric), when it expressed in different coordinates, it respond differently due to the dynamic upgradation of the constant parameters. When the metric is expressed in Schwarzschild coordinate, the dynamic upgradation of parameters lead to the change in the Ricci-scalar by a total time-derivative. However, the bulk term of each decomposition remains unchanged. If the time-dependence is obliterated introducing a smallness parameter, we obtain singular Ricci-scalar near the horizon. When the Schwarzschild metric is expressed in terms of isotropic coordinate, the change in the expression of the Ricci-scalar is not a total derivative term. However, we obtain that the Ricci-scalar still diverges upon obliteration of the time-dependence by the smallness parameter. When the Schwarzschild metric is expressed in terms of the Painleve coordinate, the bulk part of the EH Lagrangian changes (due to the dynamic upgradation) and the change of the whole EH Lagrangian can be expressed in terms of either total space- or total time-derivative. However, in this case, there is no divergence in the Ricci scalar if we remove the time dependence introducing the smallness parameter. Also, it can be shown that the divergence in the Ricci-scalar is also not present for the Kruskal coordinates.
 
 Finally, we make some comments on the divergence of the scalar curvature. Although the exact reason for the divergence of the Ricci-scalar might be revealed in the future work; nevertheless, some comments can be provided as follows. While, computing the Ricci-scalar we take derivative with respect to the proper time of a static observer at infinity. For a distant observer, the horizon is an infinite redshift surface, which implies that it takes infinite time (from the perspective of a distant observer) to see an infalling observer to cross the horizon. This may be the reason of divergence as the derivatives (such as $\dot M$) are computed with respect to the proper time of the distant observer. Moreover, the divergence in the Ricci-scalar may be a reflection of the coordinate singularity. All the metrics, for which we have obtained the divergence of the Ricci-scalar, have coordinate singularity on the horizon. In every case, when the divergence happens, the metric component which vanishes on the horizon, appears on the denominator. Therefore, one is required to explore further along this direction to determine precisely the reason for this divergence. May be, this singular behaviour might have some other significance. For example in the work of Padmanabhan \cite{Padmanabhan:2003dc}, he has correlated this phenomenon with the Bohm-Aharanov effect. We hope to report soon in this aspect to obtain more insight in this topic.

\section*{Acknowledgement}
I had started this work under the mentorship of Prof. T. Padmanabhan, who passed away after a few months after I had started to work with him. I sincerely acknowledge his guidance in this work. This work is dedicated to his memory.
 
  \vskip 3mm
 \textbf{Data availability:} This manuscript has no associated data or the data will not be deposited.

\end{document}